\documentclass[hyper]{JHEP} 

\usepackage{epsfig}

\newcommand\fverb{\setbox\pippobox=\hbox\bgroup\verb}

\newcommand\fverbdo{\egroup\medskip\noindent%

            \fbox{\unhbox\pippobox}\ }

\newcommand\fverbit{\egroup\item[\fbox{\unhbox\pippobox}]}

\newbox\pippobox

\title{Remark About  Non-Relativistic p-Brane}
\author{J. Kluso\v{n}\\
Department of
Theoretical Physics and Astrophysics\\
Faculty of Science, Masaryk University\\
Kotl\'{a}\v{r}sk\'{a} 2, 611 37, Brno\\
Czech Republic\\
E-mail: \email{klu@physics.muni.cz}} \preprint{}

 \abstract{We define different non-relativistic limit of p-brane with
 the help of canonical form of the p-brane action. We discuss properties
of these actions  and their symmetries. }

\def\tLambda{\tilde{\Lambda}}

\def\bA{\mathbf{A}}

\def\ttau{\tilde{\tau}}

\def\bB{\mathbf{B}}

\def\tx{\tilde{x}}

\def\be{\begin{equation}}

\def\ee{\end{equation}}

\def\bea{\begin{eqnarray}}

\def\eea{\end{eqnarray}}

\def\tmH{\tilde{\mH}}

\def\mH{\mathcal{H}}

\def\tG{\tilde{G}}

\def \bA{\mathbf{A}}

\newcommand{\ba}{\mathbf{a}}

\newcommand{\mL}{\mathcal{L}}

\def\pb #1{\left\{#1\right\}}

\begin{document}
\section{Introduction and Summary}
Recently new interesting formulation of non-relativistic theories was proposed in \cite{Batlle:2016iel} and further elaborated in
\cite{Gomis:2016zur,Batlle:2017cfa,Kluson:2017ufb,Kluson:2017vwp}
\footnote{For recent very nice proposal of non-relativistic string, see 
also \cite{Harmark:2017rpg}.}.
These theories belong to the class of systems with reduced symmetries that were analyzed recently from different point of views. Such a very important subject is  non-relativistic holography that is very useful tool for the study of strongly correlated systems in condensed matter, for recent review see \cite{Hartnoll:2016apf}. Non-relativistic symmetries also have fundamental meaning in the recent proposal of renormalizable quantum theory of gravity known today as Ho\v{r}ava-Lifshitz gravity \cite{Horava:2009uw}, for recent review and extensive list of references, see \cite{Wang:2017brl}. There is also an interesting connection between Ho\v{r}ava-Lifshitz gravity
and Newton-Cartan gravity \cite{Hartong:2016yrf,Hartong:2015zia}. In fact, Newton-Cartan gravity and its relation to different limits was also studied recently in series of papers
\cite{Bergshoeff:2017btm,Bergshoeff:2016lwr,
    Hartong:2015xda,Bergshoeff:2015uaa,Andringa:2010it,Bergshoeff:2015ija,Bergshoeff:2014uea}.

Another possibility how to define non-relativistic theories is to perform non-relativistic limit on the level of action for particle, string or p-brane. The first example of such object was non-relativistic string introduced in \cite{Gomis:2000bd,Danielsson:2000gi}. These actions were obtained by non-relativistic "stringy" limit where time direction and one spatial direction along the string are large. The stringy limit of superstring in $AdS_5\times S^5$ was also formulated in
\cite{Gomis:2005pg} and it was argued here that it provides another soluble sector of AdS/CFT correspondence, for related work, see \cite{Sakaguchi:2006pg,Sakaguchi:2007zsa}.  Non-relativistic limit was further extended to the case of higher dimensional objects in string theory, as for example p-branes
\cite{Gomis:2004pw,Kluson:2006xi,Brugues:2004an,Gomis:2005bj}.

It is important to stress that there is also non-relativistic limit of the relativistic string where only the time direction is large. In this case non-relativistic string does not vibrate and it represents a collection of non-relativistic massless particles.

All these limits were very carefully analyzed in \cite{Batlle:2016iel} where the general procedure how to implement non-relativistic limit for different relativistic action was proposed. The main idea is to start with an action for relativistic extended object with coordinates $X$
\begin{equation}
S=\int \mL(X)
\end{equation}
and assume that the Lagrangian density is pseudo-invariant under set of relativistic symmetries $\delta_R$
\begin{equation}\label{deltaRmL}
\delta_R \mL=\partial_\mu F^\mu \ .
\end{equation}
Then in order to find non-relativistic limit of this action we introduce dimensionless parameter $\omega$ and we define different non-relativistic limits by appropriate rescaling coordinates and parameters in Lagrangian density. Then we can presume that the Lagrangian density and symmetry transformation can be expanded in powers of $\omega$
\begin{eqnarray}
\delta_R&=&\delta_0+\omega^{-2}\delta_{-2}+\dots \ , \nonumber \\
\delta \mL&=&\omega^2\mL_2+\mL_0+\omega^{-2}\mL_{-2}+\dots \ , \nonumber \\
F^\mu&=&\omega^2 F^\mu_2+F^\mu_0+\omega^{-2}F^\mu_{-2}+\dots \ ,\nonumber \\
\end{eqnarray}
 where the first term in the expansion of the relativistic symmetry $\delta_R$ is the non-relativistic transformation $\delta_0$. Then the equation (\ref{deltaRmL}) implies infinite set of the equations when we compare expressions of the same orders in $\omega$
 \begin{eqnarray}\label{genexp}
 & &\delta_0 \mL_2=\partial_\mu F^\mu_2 \ , \nonumber \\
 & &\delta_0\mL_0+\delta_{-2}\mL_2=\partial_\mu F^\mu_0 \ , \nonumber \\
& & \delta_0 \mL_{-2}+\delta_{-2}\mL_0+\delta_{-4}\mL_2=
 \partial_\mu F^\mu_{-2} \ . \nonumber \\
 \end{eqnarray}
The special case occurs when the Lagrangian density is invariant under relativistic symmetry so that $F^\mu=0$. Then from previous equations we see that $\mL_2$ is invariant under non-relativistic symmetry while
$\mL_0$ is generally not invariant under non-relativistic symmetry.
It is further important to stress that $\mL_2$ contributes to the action with the factor $\omega^2$ and hence gives a dominant contribution in the limit $\omega \rightarrow \infty$, while $\mL_0$ remains finite and  terms proportional to $\mL_{-2},\mL_{-4},\dots$  vanish.

Since this general procedure is very interesting we mean that it is useful to explore it in more details. In particular, we would like to formulate this procedure using the canonical form of the action when we express Lagrangian density using corresponding Hamiltonian. It turns out that it is very useful since it allows us to straightforwardly identify physical degrees of freedom in the limit $\omega\rightarrow \infty$. More precisely, we introduce scaling of non-relativistic directions at the level of the action and then we find corresponding Hamiltonian for finite $\omega$. Corresponding canonical action is invariant under relativistic transformations by definitions and we also determine form of these transformations for rescaled variables for finite $\omega$. Then we discuss properties of resulting Lagrangian density in dependence on the scaling of the tension of original p-brane and on the number of non-relativistic dimensions. We argue that the non-relativistic Lagrangian $\mL_0$ is invariant under non-relativistic symmetries on condition when the Lagrangian $\mL_2$ vanish in agreement with the general discussion in \cite{Batlle:2016iel}. We also argue that in case when $\mL_2 $ is non-zero the variation of the Lagrangian density $\mL_0$ under non-relativistic transformations exactly cancels the variation $\delta_{-2}\mL_2$. On the other hand in this case we are not quite sure how to deal with divergent term in the Lagrangian which however can be canceled when we allow that p-brane couples to appropriate $p+1$-form field exactly as in
\cite{Gomis:2005bj}. However the fact that there is a background $p+1$ form
breaks the original relativistic symmetry to the subgroup that leaves this background field invariant and hence the symmetry group is reduced. More precisely, in case when we cancel divergent term the Lagrangian density $\mL_2$ is zero and hence the Lagrangian density $\mL_0$ has to be invariant under reduced group of symmetries. Of course, there is on exception  which is a fundamental string when it can be shown that in the flat space-time the Lagrangian density $\mL_2$ is total derivative and hence can be ignored \cite{Batlle:2016iel}.
We also determine Hamiltonian constraint for the non-relativistic p-brane and we show that it is linear in momenta.

As the final part of our work we focus on particle-like limit of
p-brane when only time direction is large. Using canonical form of
the action we easily find corresponding Lagrangian and we show that
it is invariant under Galilean transformations.

Let us outline our results. We propose different non-relativistic limits for p-brane when the relativistic  action has canonical form. We discuss two particular cases where the corresponding Hamiltonian constraint takes very simple form even if this procedure is completely general and serves as an analogue to the procedure suggested in \cite{Batlle:2016iel}. On the other hand using the canonical form of the action we can easily find dynamical degrees of freedom
and corresponding Hamiltonian. Then when perform inverse transformation we
derive Lagrangian density that differs from the Lagrangian density that is derived from the relativistic Lagrangian density by absence of the kinetic term for non-relativistic coordinates which is mostly seen on an example of the particle like limit of relativistic p-brane. We show that these two Lagrangian densities agree in case of fundamental string as in \cite{Batlle:2016iel}.

This paper is organized as follows. In the next section (\ref{second})
we introduce non-relativistic limit of canonical form of the action and discuss symmetries of the theory.
In section (\ref{third}) we analyze particular case when the matrix $\tG_{ij}$ is non-singular. In section (\ref{fourth}) we  discuss the possibility how to eliminate divergent term by coupling of p-brane to $p+1$ form and we derive corresponding non-relativistic Hamiltonian. In section
(\ref{fifth}) we perform particle-like non-relativistic limit of p-brane and fundamental string.
\section{ Non-Relativistic Limit of  p-Brane Canonical Action}\label{second}
In this section we formulate our proposal how to define non-relativistic p-brane using the canonical form of the action. The starting point is an  action for relativistic  p-brane
\begin{eqnarray}\label{actnongauge}
S&=&-\ttau_p\int d^{p+1}\xi
\sqrt{-\det \bA_{\alpha\beta}}
\ ,  \nonumber \\
& &\bA_{\alpha\beta}=\eta_{AB}\partial_\alpha \tx^A\partial_\beta \tx^B \ ,   \nonumber \\
\end{eqnarray}
where $\tx^A \ , A=0,\dots,d$ label embedding of  p-brane in the
target space-time and where $\eta_{AB}=\mathrm{diag}(-1,\underbrace{1\dots,1}_d)$.
It is important to stress that the action is invariant under relativistic symmetry
\begin{equation}\label{Lororig}
\tx'^A=\tLambda^A_{ \ B}\tx^B+b^A \ , \quad  \tLambda^C_{ \ A}\eta_{CD}\tLambda^D_{ \ B}=
\eta_{AB} \ ,
\end{equation}
where $\tLambda^A_{ \ B} , b^A$ are constants.
The action (\ref{actnongauge})
was the starting point for the definition of the non-relativistic
limit that  was presented in  \cite{Batlle:2016iel}. As was argued there it is possible to  define $p+1$ different non-relativistic limits
according to the number of embedding coordinates
$(0,\dots,p+1)$ that are rescaled.  Explicitly, we have
\begin{eqnarray}\label{scalx}
\tx^\mu=\omega X^\mu \ , \quad \mu=0,\dots,q \ ,
\quad
\tx^M=X^M \ , \quad  M=q+1,\dots,d \ ,\quad
\ttau=\frac{\tau}{\omega^{k_q}} \  ,
\nonumber \\
\end{eqnarray}
where the number $k_q$ depends on the form of the non-relativistic limit.
Inserting (\ref{scalx}) into definition of the matrix $\bA_{\alpha\beta}$ we obtain that
it has the form
\begin{eqnarray}\label{defbAomega}
\bA_{\alpha\beta}&=&\omega^2\tG_{\alpha\beta}+\ba_{\alpha\beta}  \ ,
\nonumber \\
\tG_{\alpha\beta}&\equiv& \partial_\alpha X^\mu \partial_\beta X_\mu \ , \quad
\ba_{\alpha\beta}=\partial_\alpha X^M\partial_\beta X_M \ .  \nonumber \\
\end{eqnarray}
Observe that we can write the matrix $\bA_{\alpha\beta}$ as
\begin{equation}
\bA_{\alpha\beta}=G_{AB}\partial_\alpha X^A\partial_\beta X^B=
G_{\mu\nu}\partial_\alpha X^\mu\partial_\beta X^\nu+
G_{MN}\partial_\alpha X^M\partial_\beta X^N \ ,
\end{equation}
where $G_{\mu\nu}=\omega^2\eta_{\mu\nu} \ , \quad
G_{MN}=\delta_{MN}$.

Our proposal is to define non-relativistic limit with the help of the
canonical form of the action.  To do this we find  Hamiltonian formalism for p-brane for finite $\omega$ and take the limit $\omega\rightarrow \infty$ after we
derive canonical form of the action.
Note that  $k_q$ is an integer number that will be determined by requirement that
there are terms in the Lagrangian density at most  quadratic at $\omega^2$.  Using (\ref{defbAomega})
we find  following conjugate momenta
\begin{eqnarray}
p_\mu&=&-\frac{\tau_p}{\omega^{k_q}}\omega^2\partial_\beta X_\mu (\bA^{-1})^{\beta 0}
\sqrt{-\det \bA} \ , \nonumber \\
p_M&=&-\frac{\tau_p}{\omega^{k_q}}
\partial_\alpha X_M (\bA^{-1})^{\alpha 0} \sqrt{-\det\bA} \ .
\end{eqnarray}
Then it is easy to see that the  bare Hamiltonian is equal to zero
\begin{eqnarray}
H_B=\int d^p\xi (p_\mu \partial_0X^\mu
+p_M\partial_0 X^M-\mL)=0
\nonumber \\
\end{eqnarray}
while we have following collection of the primary constraints
\begin{eqnarray}
\mH_i&=&p_\mu \partial_i X^\mu+p_M\partial_i X^M\approx 0 \ ,  \nonumber \\
\tmH_\tau&=&\frac{1}{\omega^2}p_\mu\eta^{\mu\nu}p_\nu+p_M p^M+\frac{\tau_p^2}{\omega^{2k_q}} \det \bA_{ij}
\approx 0 \nonumber \\
\end{eqnarray}
so that the Lagrangian density has the form
\begin{eqnarray}\label{mLomega}
\mL&=&p_\mu\partial_0 X^\mu+p_M\partial_0 X^M-
\lambda^\tau
(\frac{1}{\omega^2}p_\mu\eta^{\mu\nu}p_\nu+p_M p^M+ \frac{\tau_p^2}{\omega^{2k_q}}\det \bA_{ij})-
\nonumber \\
&-&\lambda^i (p_\mu \partial_i X^\mu+p_M\partial_i X^M) \ .
\nonumber \\
\end{eqnarray}
Let us now discuss the Lorentz transformation (\ref{Lororig})
in more details. It is instructive to write them in the form
\begin{equation}
\left(\begin{array}{cc}
\tx'^\mu \\
\tx'^M \\ \end{array}\right)=
\left(\begin{array}{cc}
\tLambda^\mu_{ \ \rho} & \tLambda^\mu_{ \ K} \\
\tLambda^M_{ \ \rho} & \tLambda^M_{ \ K} \\ \end{array}
\right)
\left(\begin{array}{cc}
\tx^\rho \\
\tx^K \\ \end{array}\right) \ .
\end{equation}
If we replace original variables with rescaled ones we obtain
\begin{eqnarray}\label{trans1}
X'^\mu&=&\tLambda^\mu_{ \ \nu}X^\nu+\frac{1}{\omega}\tLambda^\mu_{ \ M}X^M \ ,
\nonumber \\
X'^M&=&\omega \tLambda^M_{ \ \nu}X^\nu+\tLambda^M_{ \ N}X^N \ , \nonumber \\
\end{eqnarray}
where
$\tLambda^A_{ \ B}$ has to obey the equation
\begin{eqnarray}\label{tLambdarule}
\tLambda^A_{ \ C}\tLambda^B_{ \ D}\eta_{AB}=\eta_{CD}  \ .
\end{eqnarray}
It is natural to require that the  transformation rule for $X^M$  is finite in the limit $\omega\rightarrow \infty$ and hence we perform following rescaling
\begin{equation}
\tLambda^M_{ \  \nu}=\frac{1}{\omega}\Lambda^M_{ \ \nu} \ .
\end{equation}
Further, from (\ref{trans1}) we see that $\tLambda^\mu_{ \ \nu} \ ,
\tLambda^M_{ \ N}$ are not rescaled:
\begin{equation}
\tLambda^\mu_{ \ \nu}=\Lambda^\mu_{ \ \nu} \ , \quad
\tLambda^M_{ \ N}=\Lambda^M_{ \ N} \ .
\end{equation}
On the other hand if we decompose (\ref{tLambdarule}) into
corresponding components we obtain
\begin{eqnarray}
& & \Lambda^\rho_{ \ \mu}\eta_{\rho\sigma}\Lambda^\sigma_{ \ \nu}+
\frac{1}{\omega^2}\Lambda^M_{ \ \mu}\delta_{MN}\Lambda^N_{ \ \nu}=\eta_{\mu\nu} \  ,
\nonumber \\
& &\Lambda^\mu_{ \ \rho}\eta_{\mu\nu}\tLambda^\nu_{ \ M}+
\frac{1}{\omega}\Lambda^N_{ \ \rho}\delta_{NK}\Lambda^K_{ \ M}=0 \ , \nonumber \\
& &\tLambda^\mu_{ \ M}\eta_{\mu\nu}\Lambda^\nu_{ \ \rho}+
\frac{1}{\omega}\Lambda^N_{ \ M}\delta_{NK}\lambda^K_{ \ \rho}=0 \ , \nonumber \\
& &\frac{1}{\omega^2}\Lambda^\mu_{ \ M}\eta_{\mu\nu}\Lambda^\nu_{ \ N}+\Lambda^K_{ \ M}
\delta_{KL}\Lambda^L_{ \ N}=\delta_{MN} \  \nonumber \\
\end{eqnarray}
and  we see that we have to demand following scaling rule for
$\tLambda^\mu_{ \ M}$
\begin{equation}
\tLambda^\mu_{ \ N}=
 \frac{1}{\omega}\Lambda^\mu_{ \ M} \ .
\end{equation}
Using there results in  (\ref{trans1}) we obtain final form of
 Lorentz transformations for rescaled variables:
\begin{equation}
X'^\mu=\Lambda^\mu_{ \ \nu}X^\nu+\frac{1}{\omega^2}\Lambda^\mu_{ \ M}X^M \ , \quad
X'^M=\Lambda^M_{ \ N}X^N+\Lambda^M_{ \ \nu}X^\nu
\end{equation}
or in its infinitesimal form: $\Lambda^\mu_{ \ \nu}=\delta^\mu_{ \ \nu}+
\omega^\mu_{ \ \nu} \ ,  \Lambda^\mu_{\ M}=\lambda^\mu_{ \ M} \ ,
\Lambda^M_{ \ \nu}=\lambda^M_{ \ \nu}, \Lambda^M_{ \ N}=
\delta^M_{ \ N}+\omega^M_{ \ N}$
\begin{eqnarray}
\delta X^\mu&=&X'^\mu-X^\mu=\omega^\mu_{ \ \nu}X^\nu+\frac{1}{\omega^2}
\lambda^\mu_{ \ M}X^M \ ,  \nonumber \\
\delta X'^M&=&X'^M-X^M=\omega^M_{ \ N}X^N+\lambda^M_{ \ \nu}X^\nu
\nonumber \\
\end{eqnarray}
so that
\begin{eqnarray}
\delta_0 X^\mu=\omega^\mu_{ \ \nu}X^\nu \ , \quad
\delta_0 X^M=\omega^M_{ \ N}X^N+\lambda^M_{ \ \nu}X^\nu \ , \quad
\delta_{-2}X^\mu=\Lambda^\mu_{ \ M}X^M \ . \nonumber \\
\end{eqnarray}
It is important to stress that the parameters $\omega^\mu_{ \ \nu},
\lambda^M_{ \ \mu}$ can be expanded in powers of $\omega^{-2}$ so that
we obtain infinite number of terms in the expansion of the Lorentz transformations in agreement with the general definition
(\ref{genexp}). However for our purposes the number of these terms given above is sufficient.

Since we consider Lagrangian density in the canonical form we also
have to find corresponding transformation rule for conjugate
momenta. For simplicity we will consider infinitesimal form of the
transformation and we demand that   the combination $p_\mu
\partial_0X^\mu+p_M\partial_0X^M$ is invariant
\begin{equation}
\delta p_\mu \partial_0X^\mu+p_\mu \partial_0\delta X^\mu+\delta p_M \partial_0X^M+
p_M\partial_0\delta X^M=0
\end{equation}
that in the end implies following transformation rules
\begin{equation}
\delta p_\mu=-p_\nu \omega^\nu_{ \ \mu}-p_M\lambda^M_{ \ \mu} \ , \quad
\delta p_M=-p_N \omega^N_{ \ M}-\frac{1}{\omega^2}p_\mu \lambda^\mu_{ \ M} \ .
\nonumber \\
\end{equation}
It is clear that  the Lagrangian density
(\ref{mLomega}) is invariant under these transformations
since it is manifestly
 invariant under Lorentz transformations and the transformation
rules given above are ordinary Lorentz transformations rewritten with the help of the rescaled variables. Another situation occurs when
we consider specific form of the non-relativistic Lagrangian density
and study its properties in the limit $\omega \rightarrow \infty$.
%
%
It is important to stress that the Lagrangian density (\ref{mLomega})
is exact in $\omega$  and we can perform its expansion in powers of $\omega^2$ exactly as in \cite{Batlle:2016iel} even for the case when
the matrix $\tG_{ij}$ is singular.  For simplicity we restrict ourselves to two particular cases that allow to find simple result which however also describe main properties  of the procedure introduced  above. We start with the case when the matrix $\tG_{ij}$ is non-singular.
\section{The First Case: $\tG_{ij}$ is Non-singular Matrix}\label{third}
As the first possibility we consider the case when  $\tG_{ij}$ is non-singular matrix.
Note that it is $p\times p$ matrix in the form $\partial_i X^\mu \eta_{\mu\nu}
\partial_j X^\nu$ where $\partial_i X^\mu$ is $p\times (q+1)$ matrix where $q\leq p$. In case when $q+1=p$ we find that $\tG_{ij}$ is non-singular matrix and we can write
\begin{equation}
\det (
\omega^2 \tG_{ij}+\ba_{ij})=\omega^{2p}\det \tG \det (\delta_i^j
+\frac{1}{\omega^{2}}\tG^{ik}\ba_{kj}) \ .
\end{equation}
If we choose $k_q=p$ we find that the Lagrangian density has the form
\begin{eqnarray}
\mL&=&p_\mu\partial_0 X^\mu+p_M\partial_0 X^M-\nonumber \\
&-&\lambda^\tau
(\frac{1}{\omega^2}p_\mu\eta^{\mu\nu}p_\nu+p_M p^M +\tau_p^2 \det \tG_{ij}+\frac{1}{\omega^2}\tau_p^2
\det \tG_{ij}\tG^{ij}\ba_{ji})
\nonumber \\
&-&\lambda^i (p_\mu \partial_i X^\mu+p_M\partial_i X^M) \nonumber \\
\end{eqnarray}
so that we can easily take the limit $\omega\rightarrow \infty$ and we obtain
\begin{eqnarray}
\mL&=&p_\mu\partial_0 X^\mu+p_M\partial_0 X^M-\nonumber \\
&-&\lambda^\tau
(p_M p^M +\tau_p^2\det \tG_{ij})-\lambda^i (p_\mu \partial_i X^\mu+p_M\partial_i X^M) \ .  \nonumber \\
\end{eqnarray}
It is easy to see that this Lagrangian density is invariant
under transformations
\begin{eqnarray}
\delta X^\mu=\omega^\mu_{ \ \nu}X^\nu \ ,
\delta X'^M=\omega^M_{ \ N}X^N+\lambda^M_{ \ \nu}X^\nu \ , \nonumber \\
\delta p_\mu=-p_\nu \omega^\nu_{ \ \mu}-p_M\lambda^M_{ \ \mu} \ , \quad
\delta p_M=-p_N \omega^N_{ \ M}
\nonumber \\
\end{eqnarray}
using the fact that in the limit $\omega\rightarrow \infty$ we have following conditions
\begin{eqnarray}
\omega_{\rho \sigma}+\omega_{\sigma\rho}=0  \ , \quad
\omega_{KL}+\omega_{LK}=0  \ .
\nonumber \\
\end{eqnarray}
As the next step we determine canonical equations of motion  from an
extended Hamiltonian
\begin{equation}
H=\int d^p\xi (\lambda^\tau
(p_M p^M +\tau_p^2 \det \tG_{ij})\nonumber \\
+\lambda^i (p_\mu \partial_i X^\mu+p_M\partial_i X^M))
\end{equation}
so that we have following collection of the canonical equations of motion
\begin{eqnarray}
\partial_0 X^M&=&\pb{X^M,H}=2\lambda^\tau p^M+\lambda^i\partial_i X^M \ , \nonumber \\
\partial_0 p_M&=&\pb{p_M,H}=\partial_i (\lambda^i p_M) \ , \nonumber \\
\partial_0 X^\mu&=&\pb{X^\mu,H}=\lambda^ i\partial_i X^\mu \ , \nonumber \\
\partial_0 p_\mu&=&\pb{p_\mu,H}=\partial_i[2\lambda^\tau \tau_p^2
\partial_j X_\mu \tG^{ji}\det\tG_{ij}]+\partial_i[\lambda^i p_\mu] \ ,
\nonumber \\
& & p_M p^M+\tau_p^2\det \tG_{ij}=0 \ , \quad p_\mu\partial_i X^\mu+p_M
\partial_i X^M=0 \ . \nonumber \\
\end{eqnarray}
Let us try to solve these equations of motion at the spatial gauge when
$X^i=\xi^i$. Then the equation of motion for $X^i$ implies
$\lambda^i=0$ and the equations of motion for $p_M$ implies
that $p_M$ can depend on $\xi^i$ only. Further we see from the Hamiltonian constraint that the only possibility is to demand that $t$ and $X^M,p_M$ depend on $\xi^i$. Without lost of generality we presume that it depends on $\xi^1$ only and hence the matrix $\tG_{ij}$ is diagonal
in the form
\begin{equation}
\tG_{ij}=\mathrm{diag}(1-t'^2,1,\dots,1) \ .
\end{equation}
Then the equation of motion for $p_i$ are automatically satisfied for $i\neq 1$ and imply $p_i=0$ which is also in agreement with the spatial diffeomorphism constraints that imply $p_i=-p_M\partial_i X^M$. For $i=1$ the equation of motion for $p_1$ has the form
\begin{equation}
\partial_0 p_1=\partial_1[2\lambda^\tau \tau_p^2]
\end{equation}
that determines the value of the Lagrange multiplier $\lambda^\tau$
since $p_1=-p_M\partial_1 X^M$ and since $p_M$ depend on $x$ only:
\begin{eqnarray}
\nonumber \\
2\frac{\partial_1^2t}{\partial_1t}=\frac{\partial_1 \lambda^\tau}{\lambda^\tau}
\nonumber \\
\end{eqnarray}
that has the solution
\begin{equation}
 \lambda^\tau=C\partial_1t^2 \ ,
\end{equation}
where $C$ is a constant. Using this result we finally find that
$X^M=2C\partial_1t^2 p_M\xi^0+k^M$ where $k^M$ can depend on $\xi^1$ at least in principle. We see that the non-relativistic p-brane moves freely in the transverse space where however coordinates depend on $\xi^1$ through the
function $\partial_1t$. The simplest possibility os to choose $t=k\xi^1$ where $k$ has to obey the condition $k>1$. Then we can choose $\lambda^\tau=1$ for $C=\frac{1}{k^2}$ and $X^M$ has following time dependence
\begin{equation}
X^M=2p^M\xi^0 \ , p_Mp^M=\tau_p^2(k^2-1) \ .
\end{equation}

Let us now determine Lagrangian for this non-relativistic p-brane. In fact, using the equations of motion for $X^M$ and $X^\mu$ we easily find corresponding Lagrangian density
\begin{equation}
\mL=\frac{1}{4\lambda^\tau}(\partial_0 X^M-\lambda^i\partial_i X^M)
(\partial_0 X_M-\lambda^j\partial_j X_M)-\lambda^\tau\det \tG_{ij} \ .
\end{equation}
As the next step we
 eliminate Lagrange multipliers using corresponding equations of motion
\begin{eqnarray}\label{lagmuleq}
& &\frac{1}{4(\lambda^\tau)^2}(\partial_0 X^M-\lambda^i\partial_i X^M)^2
+\tau_p^2\det \tG_{ij}=0 \ , \nonumber \\
& &
 \partial_i X^M (\partial_0 X_M-\lambda^j\partial_j X_M)=0 \ .  \nonumber \\
\end{eqnarray}
To proceed further let us analyze the matrix $F_{ij}=\partial_i X^M\partial_j X_M$. This is  $p\times p$ matrix which is given as a product of  $p\times (d-q)$ matrices $\partial_i X^M$ and $(d-q)\times (d-q)$ matrix $\delta_{MN}$ and hence has the rank
$\mathrm{min}(p,(d-q))$. This matrix will be non-singular if $p<d-q$ which leads to the condition (using the fact that $q=p-1$)
\begin{equation}\label{plessd1}
p<\frac{d+1}{2} \ .
\end{equation}
 Let us presume this case and hence we can solve the last equation in
 (\ref{lagmuleq}) for $\lambda^i$ as
\begin{equation}
\lambda^i=F_{0j}F^{ji}  \ , \quad  F_{\alpha\beta}\equiv \partial_\alpha X^M\partial_\beta X_M
\end{equation}
so that the first equation in (\ref{lagmuleq}) gives
\begin{equation}
(\lambda^\tau)^2=-\frac{1}{4\tau^2\det \tG_{ij}}(F_{00}-F_{0k}F^{kj}F_{j0})=
-\frac{1}{4\tau^2\det\tG_{ij}\det F_{ij}}
\det F_{\mu\nu}
\end{equation}
and hence the Lagrangian density has the final form
\begin{equation}\label{mLF}
\mL=\tau\sqrt{-\det \tG_{ij}\frac{\det F_{\alpha\beta}}{\det F_{ij}}}
\end{equation}
Of course, the last condition holds on condition that the matrix
$F_{\alpha\beta}=\partial_\alpha X^M \partial_\beta X_M$ is non-singular. On the other hand we have that this is $(p+1)\times (p+1)$ matrix
with the rank given as $\mathrm{min}(p+1,d-(p-1))$ that implies
\begin{equation}
p<\frac{d}{2} \
\end{equation}
that is stronger condition than the condition
(\ref{plessd1}) but certainly can be obeyed.
Finally note that (\ref{mLF}) is invariant under
 scaling transformations
\begin{equation}
X'^\mu=\lambda^{-\frac{1}{p}}X^\mu \ , \quad X'^M=\lambda X^M \ .
\end{equation}
It is interesting  that the Lagrangian density
derived from the Hamiltonian is not unique. For example, let us consider an equation of motion for $X^\mu$
\begin{equation}
\partial_0 X^\mu=\lambda^i\partial_i X^\mu \ .
\end{equation}
Multiply this equation with $\partial_j X_\mu$ we obtain
\begin{equation}
\partial_0 X^\mu\partial_j X_\mu=\lambda^i \tG_{ij}
\end{equation}
that can be solved for $\lambda^i$ using the fact that $\tG_{ij}$ is
non-singular
\begin{equation}
\lambda^i=\tG^{ij}\tG_{j0} \ , \quad \tG_{\alpha\beta}=
\partial_\alpha X^\mu \partial_\beta X_\nu \ .
\end{equation}
Then  the equation of motion for $\lambda^\tau$ has the form
\begin{eqnarray}
(\lambda^\tau)^2=-\frac{1}{4\tau^2 \det\tG_{ij}}
(\ba_{00}-2\tG_{0j}\tG^{ji}\ba_{i0}+\tG_{0i}\tG^{ik}\ba_{kl}
\tG^{lj}\tG_{j0})
\end{eqnarray}
and hence we obtain following  Lagrangian density
\begin{equation}
\mL=-\tau \sqrt{(\ba_{00}-2\tG_{0j}\tG^{ji}\ba_{j0}+
    \tG_{0i}\tG^{ik}\ba_{kl}\tG^{lj}\tG_{j0})\det\tG_{ij}} \ .
\end{equation}

We argued that the form of non-relativistic Lagrangian density depends
on the value of the coefficient $k_q$. Let us  consider the  second possibility when
$k_q$ is equal to $k_q=p-1$ so that the Lagrangian density is equal to
\begin{eqnarray}
\mL&=&p_\mu\partial_0 X^\mu+p_M\partial_0 X^M-\nonumber \\
&-&\lambda^\tau
(\frac{1}{\omega^2}p_\mu\eta^{\mu\nu}p_\nu+p_M p^M +\omega^2\tau_p^2 \det \tG_{ij}+\tau_p^2
\det \tG_{ij}\tG^{ij}\ba_{ji})
\nonumber \\
&-&\lambda^i (p_\mu \partial_i X^\mu+p_M\partial_i X^M) \nonumber \\
\end{eqnarray}
so that we can easily take the limit $\omega\rightarrow \infty$  and we obtain
\begin{eqnarray}
\mL&=&\omega^2 \mL_2+\mL_0  \ , \quad
\mL_2=-\lambda^\tau \tau_p^2\det \tG_{ij} \ , \nonumber \\
\mL_0&=&p_\mu\partial_0 X^\mu+p_M\partial_0 X^M-\nonumber \\
&-&\lambda^\tau
(p_M p^M +\tau_p^2\det \tG_{ij}\tG^{ij}\ba_{ij})-\lambda^i (p_\mu \partial_i X^\mu+p_M\partial_i X^M) \ .  \nonumber \\
\end{eqnarray}
Now we would like to analyze transformation rules for different terms in the Lagrangian. Clearly we have $\delta_{0}\mL_2=0$ while we have
\begin{eqnarray}
\delta_0 \mL_0&=&-\tau_p^2 \det \tG_{ij}\delta \ba_{ij}=\nonumber \\
&=&-\tau_p^2\det \tG_{ij}\tG^{ij}(\lambda^M_{ \ \nu}\partial_i X^\nu
\delta_{MN}\partial_j X^N+\partial_i X^M\delta_{MN}\lambda^N_{ \ \nu}
\partial_j X^\nu) \ .  \nonumber \\
\end{eqnarray}
On the other hand it is easy to see that
\begin{eqnarray}
\delta_{-2}\mL_2&=&-\tau_p^2\det \tG_{kl}\delta_{(-2)} \tG_{ij}\tG^{ji}=
\nonumber \\
&=&-\tau_p^2\det \tG_{kl}\tG^{ij}(\partial_i (\lambda^\mu_{ \ M}X^M)
\eta_{\mu\nu}\partial_j X^\nu+\partial_i X^\mu \eta_{\mu\nu}
\partial_j (\lambda^\nu_{ \ M}X^M)) \nonumber \\
\end{eqnarray}
using the fact that $\delta_{-2}X^\mu=\Lambda^\mu_{ \ M}X^M$.
Now it is easy to see that $\delta_{-2}\mL_2+\delta_0\mL_0=0$
thanks to the conditions
\begin{eqnarray}
\lambda^\mu_{ \ M}\eta_{\mu\rho}+\delta_{MK}\lambda^K_{ \ \rho}=0 \ , \quad
\eta_{\rho\nu}\lambda^\nu_{ \ M}+\lambda^K_{ \ \rho}\delta_{KM}=0 \ .
\nonumber \\
\end{eqnarray}
  It is important to stress that the variation of the Lagrangian density $\mL_0$ proportional to $\lambda^M_{ \ \mu}$ is compensated by the variation of the Lagrangian density $\mL_2$ which is in agreement with \cite{Batlle:2016iel}. On the other hand it is not completely clear how to deal  the presence of the divergent term in the Lagrangian when we analyze corresponding equations of motion. The well defined procedure how to eliminate this term is to couple p-brane to background $p+1$ form.
  We will show that this can be done in  canonical approach too.
\section{Elimination of Divergent Term}\label{fourth}
We begin this section with the case of the massive relativistic particle action in the form
\begin{equation}
S=\int d\tau (p_t \dot{t}+p_M\dot{X}^M-
e(-\frac{1}{\omega^2}p_t^2+p_M^2+\tilde{m}^2))
\end{equation}
and we see that the limit $\omega\rightarrow
\infty$ gives the result
\begin{equation}
S=\int d\tau (p_t \dot{t}+p_M\dot{X}^M-
e(p_M^2+\tilde{m}^2)) \ .
\end{equation}
From the Hamiltonian constraint $p_M^2+m^2=0$ we see that
the non-relativistic limit can be defined if we scale $\tilde{m}^2=
\frac{1}{\omega^2}m^2$. On the other hand the dynamics is still trivial since the Hamiltonian constraint implies  $p_Mp^M=0$
that has solution $p_M=0$. In order to resolve this
problem let us consider the possibility that we couple the particle to the background
electromagnetic field so that the action has the form
\begin{equation}
S=-\tilde{m}\int d\tau \sqrt{-g_{AB}\dot{X}^A\dot{X}^B}+M\int d\tau A_A\dot{X}^A
\end{equation}
so that the conjugate momentum is
\begin{equation}
p_A=\frac{\tilde{m} g_{AB}\dot{X}^B}{\sqrt{-g_{AB}\dot{X}^A\dot{X}^B}}+MA_A
\end{equation}
that implies following constraint
\begin{equation}
\mH_\tau =(p_A-MA_A)g^{AB}(p_B-MA_B)+\tilde{m}^2\approx 0 \ .
\end{equation}
We define non-relativistic limit when $g_{AB}=\mathrm{diag}(-\omega^2,1,\dots,1),
g^{AB}=\mathrm{diag}(-\frac{1}{\omega^2},1,\dots,1)$ and also
$A_0=\omega^2$. Then the Hamiltonian constraint has the form
\begin{equation}
\mH_\tau=p_M p^M-\frac{1}{\omega^2}p_t^2+2Mp_t-M^2\omega^2+\tilde{m}^2\approx 0
\end{equation}
Now we see that we derive well defined limit when we scale $\tilde{m}^2$ as
$\tilde{m}^2=M^2\omega^2$ so that the Hamiltonian for non-relativistic particle has
the form
\begin{equation}
\mH_\tau=p_M p^M+2Mp_t \approx 0  \ .
\end{equation}
This is clearly non-relativistic Hamiltonian constraint and we see that
it was crucial that the particle coupled to the gauge field. It is now easy
to determine corresponding Lagrangian density using the equation of motion
for $X^M$ and for $t$
\begin{equation}
\dot{X}^M=\pb{X^M,H}=2ep_M \ , \quad
\dot{t}=\pb{t,H}=2Me
\end{equation}
and hence
\begin{equation}
L=p_M\dot{X}^M+p_t\dot{t}-H=
ep_M p^M=\frac{1}{4e}\dot{X}^M\dot{X}_M=\frac{M}{2\dot{t}}
\dot{X}^M\dot{X}_M \ ,
\end{equation}
where in the last step we used equation of motion for $t$. Note that this Lagrangian has the same form as the Lagrangian for non-relativistic particle derived in \cite{Andringa:2012uz}.

As the next step we consider  fundamental string coupled to background NSNS two form where the action has the form
\begin{eqnarray}
S=-\ttau \int d\tau d\sigma \sqrt{-\det g_{\alpha\beta}}
+\ttau\int d\tau d\sigma B_{AB}\partial_\tau X^A\partial_\sigma X^B
\nonumber \\
\end{eqnarray}
so that we have following conjugate momenta
\begin{equation}
p_A=-\ttau G_{AB}\partial_\alpha X^B g^{\beta\tau}\sqrt{-\det g_{\alpha\beta}}+
\ttau B_{AB}\partial_\sigma X^B
\end{equation}
that implies following Hamiltonian constraint
\begin{equation}
\mH_\tau=
(p_A+\ttau B_{AC}\partial_\sigma X^C)g^{AB}(p_B+\ttau B_{BD}\partial_\sigma X^D)+\ttau^2 G_{AB}\partial_\sigma X^A
\partial_\sigma X^B\approx 0 \ .
\end{equation}
In order to define stringy  non-relativistic limit
we choose following components of the metric $G_{\mu\nu}=\omega^2
\eta_{\mu\nu} \ , G^{\mu\nu}=\frac{1}{\omega^2}\eta^{\mu\nu} \ , \mu,\nu=0,1 $ so that we obtain Hamiltonian constraint in the form
\begin{eqnarray}
\mH_\tau&=&\frac{1}{\omega^2}p_\mu \eta^{\mu\nu}p_\nu+
2p_\mu\ttau \frac{1}{\omega^2}\eta^{\mu\nu}B_{\nu\sigma}\partial_\sigma X^\sigma+p_M g^{MN}p_N+\nonumber \\
&+&\frac{1}{\omega^2}\tau^2 B_{\mu\rho}\partial_\sigma X^\rho \eta^{\mu\nu}
B_{\nu\sigma}\partial_\sigma X^\sigma+\ttau^2 \omega^2
\eta_{\mu\nu}\partial_\sigma X^\mu \partial_\sigma X^\nu+
\ttau^2 \partial_\sigma X^M\partial_\sigma X_M
\approx 0 \ .
\nonumber \\
\end{eqnarray}
We see that the divergent term can be eliminated by suitable choice of the background NSNS two form when we take
 $B_{\mu\nu}=\omega^2 \epsilon_{\mu\nu} \ ,
\epsilon_{01}=-1$. Further, the string tension is not rescaled
$\ttau=\tau$
 and hence the Hamiltonian constraint has the form
\begin{equation}
\mH_\tau=2\tau p_\mu \eta^{\mu\nu}\epsilon_{\nu\sigma}\partial_\sigma X^\sigma+p_Mp^M+\tau^2 \partial_\sigma X^M\partial_\sigma X_M\approx 0 \ .
\end{equation}
This is the same form of the Hamiltonian constraint as was derived in
\cite{Gomis:2004ht}.

The generalization of this procedure to the case of $p-$brane is
straightforward. We presume that this $p-$brane couples to
 $C^{p+1}$ form
so that the action has the form
\begin{equation}
S=-\ttau_p\int d^{p+1}\xi\sqrt{-\det \bA_{\alpha\beta}}
+\ttau_p\int C^{(p+1)} \ ,
\end{equation}
where
\begin{equation}
C^{(p+1)}=C_{A_1\dots A_{p+1}}dX^{A_1}\wedge \dots dX^{A_{p+1}}=
\frac{1}{(p+1)!}\epsilon^{\alpha_1\dots \alpha_{p+1}}
C_{A_1\dots A_{p+1}}\partial_{\alpha_1}X^{A_1}\dots \partial_{\alpha_{p+1}}X^{A_{p+1}}
\end{equation}
so that we have following conjugate momenta
\begin{equation}
p_A=-\ttau G_{AB}\partial_\beta X^B(\bA^{-1})^{\beta 0}
\sqrt{-\det \bA}+\frac{\ttau_p}{p!}C_{AA_2\dots A_{p+1}}
\epsilon^{i_2\dots i_{p+1}}\partial_{i_2}X^{A_2}\dots
\partial_{i_{p+1}}X^{A_{p+1}} \ .
\end{equation}
Then it is easy to see that the  Hamiltonian constraint has
the form
\begin{eqnarray}
\mH_\tau&=&(p_A-\frac{\ttau_p}{p!}C_{AA_2\dots A_{p+1}}
\epsilon^{i_2\dots i_{p+1}}\partial_{i_2}X^{A_2}\dots
\partial_{i_{p+1}}X^{A_{p+1}})g^{AB}\times \nonumber \\
&\times &(p_B-\frac{\ttau_p}{p!}C_{BB_2\dots B_{p+1}}
\epsilon^{j_2\dots j_{p+1}}\partial_{j_2}X^{B_2}\dots
\partial_{j_{p+1}}X^{B_{p+1}})+\ttau_p^2 \det \bA_{ij}\approx 0 \ .
\nonumber \\
\end{eqnarray}
Now we presume that the metric has the form $
G_{\mu\nu}=\omega^2 \eta_{\mu\nu} \ , G_{MN}=\delta_{MN} \ ,
G^{\mu\nu}=\frac{1}{\omega^2}\eta^{\mu\nu} \ , \mu,\nu=0,\dots,p$
so that we have
\begin{equation}
\det \bA_{ij}=\omega^{2p}\det \tG_{ij}+\omega^{2p-2}\det \tG_{ij}
\tG^{kl}\ba_{kl}
\end{equation}
and hence  we obtain finite result if
$\ttau_p^2 \omega^{2(p-1)}=\tau_p$ and if we choose components of the
$p+1$ form along $0,\dots,p$ directions in the form
\begin{equation}\label{Cback}
C_{\mu_0\dots\mu_{p+1}}=\omega^{p+1}\epsilon_{\mu_0\mu_1\dots\mu_{p+1}} \ .
\end{equation}
To proceed further we use the fact that
\begin{eqnarray}
& &\frac{\ttau_p^2\omega^{2p}}{(p!)^2}
C_{\mu\mu_2\dots\mu_{p+1}}\epsilon^{i_2\dots i_{p+1}}
\partial_{i_2}X^{\mu_2}\dots\partial_{i_{p+1}}X^{\mu_{p+1}}
\eta^{\mu\nu}
C_{\nu\nu_2\dots\nu_{p+1}}\epsilon^{j_2\dots j_{p+1}}
\partial_{j_2}X^{\nu_2}\dots\partial_{j_{p+1}}X^{\nu_{p+1}}=
\nonumber \\
&=&-\frac{\ttau_p^2\omega^{2p}}{p!}\epsilon^{i_1\dots i_{p}}\epsilon^{j_1\dots j_{p}}
\partial_{i_1}X^{\mu_1}\partial_{j_1}X_{\mu_1}\dots
\tG_{i_1j_1}\dots \tG_{i_pj_p}=
-\ttau_p^2\omega^{2p}\det \tG_{ij} \nonumber \\
\end{eqnarray}
and we see that these two divergent contributions cancel. In other words we have following
final form of the Hamiltonian constraint
\begin{equation}\label{mHtaup}
\mH_\tau=-2\frac{\tau_p}{p!}p^\mu\epsilon_{\mu\mu_2\dots\mu_{p+1}}
\epsilon^{i_2\dots i_{p+1}}\partial_{i_2}X^{\mu_2}
\dots\partial_{i_{p+1}}X^{\mu_{p+1}}+p_M p^M+\tau_p^2
\det \tG_{ij}\tG^{kl}\ba_{kl}\approx 0
\end{equation}
and we see that this Hamiltonian constraint is linear in non-relativistic  momenta. Note that the Hamiltonian constraint is invariant under
transformations
\begin{equation}\label{Lorsubgroup}
\delta X^\mu=\omega^\mu_{ \ \nu}X^\nu \ , \quad
\delta X^M=\omega^M_{ \ N}X^N
\ ,
\end{equation}
where we observe an important fact that there is absent the mixed
term $\lambda^M_{ \ \mu}X^\mu$ in the variation of $X^M$. This is a consequence of the fact that the presence of the background $p+1$ form
breaks the original Lorentz symmetry to the transformations
(\ref{Lorsubgroup}) that leaves the background $p+1$ form
(\ref{Cback}) invariant.

It is instructive to compare the Hamiltonian constraint (\ref{mHtaup})
wit the one that was derived in \cite{Kluson:2017vwp} where the Hamiltonian analysis
of non-BPS Dp-brane was performed. The Hamiltonian constraint derived
in \cite{Kluson:2017vwp} can be easily truncated to the case of p-brane and we obtain
\begin{equation}
\mH_{\tau}^{sq.r.}=
p_M p^M+\tau_p^2
\det \tG_{ij}\tG^{kl}\ba_{kl}-\tau_p
\sqrt{-p_\mu (\eta^{\mu\nu}-\partial_i X^\mu \tG^{ij}
    \partial_j X^\nu)p_\nu} \ .
\end{equation}
Since we are interested in
the physical content of the theory it is natural to consider the
gauge fixed theory and hence we impose spatial static gauge
\begin{equation}
X^i=\xi^i
\end{equation}
so that $\tG_{ij}=\delta_{ij}$ and hence (\ref{mHtaup}) is equal to
\begin{eqnarray}
\mH_\tau
=-2\tau_p p_0+p_M p^M+\tau_p^2 \delta^{ij}\ba_{ij}\approx 0 \ . \nonumber \\
\end{eqnarray}
that agree with $\mH_\tau^{sq.r}$ evaluated at the spatial static gauge too and hence we see that  these two Hamiltonian constraints are equivalent.
\section{Particle Limit of p-Brane Action}\label{fifth}
Finally we consider the case when  we perform particle like non-relativistic limit of
p-brane action where only the time direction is large
\begin{eqnarray}\label{partlimit}
\tx^0&=&\omega t \ , \quad \tx^M= X^M \ , \quad M=1,\dots,d \ , \nonumber \\
\ttau_p&=&\frac{1}{\omega} \tau_p \ . \quad
\end{eqnarray}
In this case we have following  primary constraints
\begin{eqnarray}
\mH_i&=&p_t\partial_i t+p_M\partial_i X^M
=0\nonumber \\
\mH_\tau&=&-\frac{1}{\omega^2}p_t^2+p_M p^M+\frac{\tau_p^2}{\omega^2}
\det \bA_{ij} \approx 0  \ ,  \nonumber \\
\end{eqnarray}
where $\bA_{ij}=-\omega^2 \partial_i t\partial_j t+\ba_{ij}$.
Now we can write
\begin{eqnarray}
\det \bA_{ij}&=&\frac{1}{p!}\epsilon^{i_1\dots i_p}\epsilon^{j_1\dots j_p}
(-\omega^2 \partial_{i_1}t\partial_{j_1}t+\ba_{i_1 j_1})\times
\dots (-\omega^2 \partial_{i_p}t\partial_{j_p}t+\ba_{i_p j_p})=\nonumber \\
&=&-\omega^2\frac{1}{(p-1)!}\epsilon^{i_1\dots i_p}\epsilon^{j_1\dots j_p}
\partial_{i_1}t\partial_{j_1}t\ba_{i_2 j_2}\dots \ba_{i_pj_p}+
\det \ba \ ,  \nonumber \\
\end{eqnarray}
where all terms of higher order in $\omega^2$ vanish due to the antisymmetry of $\epsilon^{i_1\dots i_p}$. Then it is easy to see that
 the Hamiltonian constraint has the form
\begin{eqnarray}
\mH_\tau=-\frac{1}{\omega^2}p_t^2+p_M p^M-
\frac{\tau_p^2}{(p-1)!}\epsilon^{i_1\dots i_p}\epsilon^{j_1\dots j_p}
\partial_{i_1}t\partial_{j_1}t\ba_{i_2 j_2}\dots \ba_{i_pj_p}
+\frac{\tau_p^2}{\omega^2}\det\ba \approx 0   \nonumber \\
\end{eqnarray}
and we see that it is well defined for $\omega\rightarrow \infty$ when we obtain
\begin{equation}
\mH_\tau=p_M p^M-
\frac{\tau_p^2}{(p-1)!}\epsilon^{i_1\dots i_p}\epsilon^{j_1\dots j_p}
\partial_{i_1}t\partial_{j_1}t\ba_{i_2 j_2}\dots \ba_{i_pj_p} \approx 0 \ .
\end{equation}
It is easy to see that the Hamiltonian constraint is invariant under non-relativistic transformations
\begin{equation}
\delta p_M=-\omega_M^{ \ N}p_N \ , \quad \delta X^M=\omega^M_{ \ N}X^N+
\lambda^M t
\end{equation}
using the fact that  $\delta a_{ij}=\lambda^M(\partial_i t\partial_j X_M+
\partial_i X_M\partial_j t)$  and then using an antisymmetry of
$\epsilon^{i_1\dots i_p}$.

It is also instructive to determine corresponding Lagrangian. Note that
 the total Hamiltonian has the form
\begin{equation}
H=\int d^p\xi (\lambda^\tau \mH_\tau+\lambda^i\mH_i)
\end{equation}
so that we have following equations of motion
\begin{eqnarray}
\partial_0 X^M=\pb{X^M,H}=2\lambda^\tau p^M+\lambda^i\partial_i X^N  \ ,  \quad
\partial_ 0 t=\pb{t,H}=\lambda^i\partial_i  t
\nonumber \\
\end{eqnarray}
and hence the Lagrangian density has the form
\begin{eqnarray}
\mL&=&p_M\partial_0 X^M+p_t\partial_0 t-\lambda^\tau \mH_\tau-
\lambda^i\mH_i=\nonumber \\
&=&\frac{1}{4\lambda^\tau}
(\ba_{00}-2\lambda^i\ba_{i0}+\lambda^i\lambda^j \ba_{ij})
+
\lambda^\tau \frac{\tau_p^2}{(p-1)!}\epsilon^{i_1\dots i_p}\epsilon^{j_1\dots j_p}
\partial_{i_1}t\partial_{j_1}t\ba_{i_2 j_2}\dots \ba_{i_pj_p} \ ,
\nonumber \\
\end{eqnarray}
where again $\ba_{\alpha\beta}=\partial_\alpha X^M\partial_\beta X_M$.
To proceed further we
 solve the equations of motion for $\lambda^i$ and $\lambda^\tau$
\begin{eqnarray}
& & \ba_{i0}-\ba_{ij}\lambda^j=0 \ , \nonumber \\
& & -\frac{1}{4(\lambda^\tau)^2}
 (\ba_{00}-2\lambda^i\ba_{i0}+\lambda^i\lambda^j \ba_{ij})+
\frac{\tau_p^2}{(p-1)!}\epsilon^{i_1\dots i_p}\epsilon^{j_1\dots j_p}
\partial_{i_1}t\partial_{j_1}t\ba_{i_2 j_2}\dots \ba_{i_pj_p} =0 \ .
\nonumber \\
\end{eqnarray}
If we presume that $\ba_{ij}$ has an inverse we can find solution of the first equation as
\begin{equation}
\lambda^i=\ba^{ij}\ba_{j0} \
\end{equation}
so that the equation of motion for $\lambda^\tau$ has the form
\begin{eqnarray}
-\frac{1}{4(\lambda^\tau)^2}
\frac{\det \ba_{\alpha\beta}}{\det \ba_{ij}}+
+
\frac{\tau_p^2}{(p-1)!}\epsilon^{i_1\dots i_p}\epsilon^{j_1\dots j_p}
\partial_{i_1}t\partial_{j_1}t\ba_{i_2 j_2}\dots \ba_{i_pj_p}=0 \nonumber \\
\end{eqnarray}
and hence the Lagrangian density has the form
\begin{equation}
\mL=\tau_p\sqrt{\frac{\det\ba_{\alpha\beta}}{\det\ba_{ij}(p-1)!}
    \epsilon^{i_1\dots i_p}\epsilon^{j_1\dots j_p}
    \partial_{i_1}t\partial_{j_1}t\ba_{i_2 j_2}\dots \ba_{i_pj_p}}
\end{equation}
It is clear that
this analysis is valid  for $p>1$. The case $p=1$ will be studied
separately in the next subsection.
\subsection{The Case of Fundamental String}
In this case the Hamiltonian constraint has the form
\begin{eqnarray}
\mH_\tau=-\frac{1}{\omega^2}p_t^2+p_M p^M
-\tau_F^2 (\partial_\sigma t\partial_\sigma t-\frac{1}{\omega^2}\partial_\sigma X^M
\partial_\sigma X_M) \approx 0
\nonumber \\
\end{eqnarray}
that implies in the limit $\omega \rightarrow \infty$
following Hamiltonian constraint
\begin{equation}\label{mHtauparticle}
\mH_\tau=
p_M p^M-\tau_F^2 \partial_\sigma t\partial_\sigma t \approx 0 \nonumber \\
\end{equation}
which agrees with the Hamiltonian constraints found in
\cite{Batlle:2016iel}.

It is interesting to find corresponding Lagrangian density. To do this we use again canonical equations of motion
\begin{eqnarray}
\partial_\tau X^M&=&\pb{X^M,H}=2\lambda^\tau p_M+\lambda^\sigma \partial_\sigma X^M \ ,
\nonumber \\
\partial_\tau t&=&\pb{t,H}=\lambda^\sigma \partial_\sigma t
\nonumber \\
\end{eqnarray}
and hence the Lagrangian density has the form
\begin{eqnarray}
\mL&=&p_M\partial_\tau X^M+p_t\partial_\tau t-\lambda^\tau \mH_\tau-
\lambda^\sigma \mH_\sigma=
\nonumber \\
&=&\frac{1}{4\lambda^\tau}(\partial_\tau X^M-\lambda^\sigma \partial_\sigma X^M)
(\partial_\tau X_M-\lambda^\sigma \partial_\sigma X_M)+\lambda^\tau \tau_F^2
\partial_\sigma t\partial_\sigma t \ .  \nonumber \\
\end{eqnarray}
Solving the equation of motion for $\lambda^\sigma$ we obtain
\begin{equation}
\lambda^\sigma=\frac{\ba_{\tau\sigma}}{\ba_{\sigma\sigma}}
\end{equation}
while the equation of motion for $\lambda^\tau$ has the form
\begin{equation}
-\frac{1}{4(\lambda^\tau)^2}
\left(\ba_{\tau\tau}-\frac{\ba_{\tau\sigma}^2}{\ba_{\sigma\sigma}}\right)
+\tau^2_F\partial_\sigma t\partial_\sigma t =0 \ .
\end{equation}
Inserting this result into original Lagrangian density we finally obtain
\begin{equation}
\mL=
\tau_F\sqrt{\frac{\det\ba_{\alpha\beta}}{\ba_{\sigma\sigma}}\partial_\sigma
    t\partial_\sigma t}  \ .\nonumber \\
\end{equation}
It is interesting that this Lagrangian does not have the same form as the
Lagrangian found in \cite{Batlle:2016iel}. Explicitly, the Lagrangian
density derived there has the form
\begin{eqnarray}\label{Lorg}
\mL&=&-\tau_F\int d\tau d\sigma
\sqrt{(\partial_\tau t\partial_\sigma X^M-\partial_\sigma t
\partial_\tau X^M)(\partial_\tau t \partial_\sigma X_M-\partial_\sigma t
\partial_\tau X_M)}=\nonumber \\
&=&-\tau_F\int d\tau d\sigma
\sqrt{\partial_\tau t\partial_\tau t \ba_{\sigma\sigma}-
2\partial_\sigma t\partial_\tau t\ba_{\tau\sigma}+
\partial_\sigma t\partial_\sigma t \ba_{\tau\tau}}=-\tau_F
\int d\tau d\sigma \sqrt{\bB} \ .
\nonumber \\
\end{eqnarray}
From (\ref{Lorg}) we obtain following conjugate momenta
\begin{eqnarray}
p_t&=&-\tau_F \frac{\partial_\tau t \ba_{\sigma\sigma}-\partial_\sigma  t
\ba_{\tau\sigma}}
{\sqrt{\bB}} \ ,
\nonumber \\
p_M&=&-\tau_F\frac{\partial_\tau X_M\partial_\sigma t
\partial_\sigma t-\partial_\sigma t\partial_\tau t \partial_\sigma X_M}
{\sqrt{\bB}} \ .
\nonumber \\
\end{eqnarray}
We again find that the bare Hamiltonian is zero
while we have following collection of the primary constraints
\begin{eqnarray}
\mH_\sigma&=&
p_M\partial_\sigma X^M+p_t\partial_\sigma t \approx 0  \ , \nonumber \\
\mH_\tau&=&p_Mp^M-\tau_F^2 \partial_\sigma t\partial_\sigma t\approx 0 \ .
\nonumber \\
\end{eqnarray}
It is interesting that in case of fundamental string we can find the same form of the Lagrangian density as in \cite{Batlle:2016iel}. To see this note that the equation of motion for $t$ implies
\begin{equation}
\lambda^\sigma=\frac{\partial_\tau t}{\partial_\sigma t}
\end{equation}
so that the Lagrangian has the form
\begin{equation}\label{mLtt}
\mL=\frac{1}{4\lambda^\tau}\left(\ba_{\tau\tau}-2\frac{\partial_\tau t}{\partial_\sigma t}\ba_{\tau\sigma}+\frac{\partial_\tau t\partial_\tau t}
{\partial_\sigma t\partial_\sigma t}\ba_{\sigma\sigma}\right)+\lambda^\tau
\tau_F^2\partial_\sigma t\partial_\sigma t \ .
\end{equation}
Then solving the equation of motion for $\lambda^\tau$ we find
\begin{equation}
\lambda^\tau=-\frac{1}{2\tau_F}
\sqrt{\frac{\ba_{\tau\tau}-2\frac{\partial_\tau t}{\partial_\sigma t}\ba_{\tau\sigma}+\frac{\partial_\tau t\partial_\tau t}{\partial_\sigma t
        \partial_\sigma t}\ba_{\sigma\sigma}}{\partial_\sigma t\partial_\sigma t}} \ .
\end{equation}
Inserting this result back into (\ref{mLtt}) we finally obtain
\begin{equation}
\mL=-\tau_F\sqrt{\ba_{\tau\tau}\partial_\sigma t\partial_\sigma t-
2\partial_\tau t\partial_\sigma t \ba_{\tau\sigma}+\partial_\tau t
\partial_\tau t\ba_{\sigma\sigma}}
\end{equation}
that agrees with (\ref{Lorg}).

\acknowledgments{This  work  was
    supported by the Grant Agency of the Czech Republic under the grant
    P201/12/G028. }

\end{document}